\documentclass{article}
\usepackage{latexsym,deluxetable}

\begin{document}

\begin{center}
IMPROVED LABORATORY TRANSITION PROBABILITIES FOR Gd \textsc{ii} AND 
APPLICATION TO THE GADOLINIUM ABUNDANCES OF THE SUN AND THREE $r-$PROCESS RICH, 
METAL-POOR STARS
\end{center}

\begin{center}
(short title: Gd Transition Probabilities and Abundances)
\end{center}

\begin{center}
E. A. Den Hartog, J. E. Lawler
\end{center}

\begin{center}
Department of Physics, University of Wisconsin, Madison, WI 53706;
\end{center}

\begin{center}
eadenhar@wisc.edu, jelawler@wisc.edu
\end{center}

\begin{center}
C. Sneden
\end{center}

\begin{center}
Department of Astronomy and McDonald Observatory, University of Texas, 
Austin, TX 78712; chris@verdi.as.utexas.edu
\end{center}

\begin{center}
and J. J. Cowan
\end{center}

\begin{center}
Department of Physics and Astronomy, University of Oklahoma, Norman, OK 
73019; cowan@nhn.ou.edu
\end{center}

\begin{center}
ABSTRACT
\end{center}
Radiative lifetimes, accurate to $\pm $5{\%}, have been measured for 49 
even-parity and 14 odd-parity levels of Gd \textsc{ii} using laser-induced 
fluorescence. The lifetimes are combined with branching fractions measured 
using Fourier transform spectrometry to determine transition probabilities 
for 611 lines of Gd \textsc{ii}. This work is the largest-scale laboratory 
study to date of Gd \textsc{ii} transition probabilities and the first using 
a high performance Fourier transform spectrometer. This improved data set 
has been used to determine a new solar photospheric Gd abundance, log 
$\varepsilon $ = 1.11 $\pm $ 0.03. Revised Gd abundances have also been 
derived for the $r-$process-rich metal-poor giant stars CS 22892-052, 
BD+17$^{o}$3248, and HD 115444. The resulting Gd/Eu abundance ratios are in 
very good agreement with the solar-system $r-$process ratio. We have employed 
the increasingly accurate stellar abundance determinations, resulting in 
large part from the more precise laboratory atomic data, to predict directly 
the Solar System $r-$process elemental abundances for Gd, Sm, Ho and Nd. Our 
analysis of the stellar data suggests slightly higher recommended values for 
the $r-$process contribution and total Solar System values, consistent with the 
photospheric determinations, for the elements for Gd, Sm, and Ho.

\bigskip
Subject headings: atomic data --- stars: abundances stars: Population II --- 
Sun: abundances

\newpage 
\begin{center}
1. INTRODUCTION
\end{center}

Rare Earth (RE) elements are among the most spectroscopically accessible of 
the neutron ($n$-) capture elements. Many transitions of singly ionized RE 
species appear in the spectrum of the Sun and in stars over a significant 
temperature range. This accessibility makes RE species useful in studies of 
heavy element nucleosynthesis. The needed observations in the form of 
high-resolution, high signal-to-noise (S/N) spectra on a variety of targets 
from very large ground-based telescopes and the Hubble Space Telescope are 
available. Old metal-poor Galactic halo stars are very attractive targets 
because they provide a fossil record of the chemical make-up of our Galaxy 
when it, and the Universe, were very young (e.g., Gratton {\&} Sneden 1994; 
McWilliam et al. 1995; Cowan et al. 1995; Sneden et al. 1996, Ryan et al. 
1996, Cayrel et al. 2004). Recent abundance determinations of heavy 
$n$-capture elements in very metal-poor stars have yielded new insights on the 
roles of the $r$(apid)- and $s$(low)-processes in the initial burst of Galactic 
nucleosynthesis. The results of this ongoing work are reshaping our 
understanding of the chemical evolution of the Galaxy.

Although these studies of nucleosynthesis are proceeding at a good pace, 
there is a constant need for improved laboratory data, especially atomic 
transition probabilities. The accessibility of RE spectra from an 
observational astronomer's viewpoint is not matched by tractability from a 
theoretical atomic physicist's viewpoint. The combination of an open 
f-shell with hundreds to thousands of low lying levels, a breakdown of 
Russell-Saunders coupling, substantial relativistic effects, and massive 
configuration interactions makes the calculation of \textit{ab initio} atomic transition 
probabilities a formidable undertaking. The Li\'{e}ge group (e.g. 
Bi\'{e}mont {\&} Quinet (2003)) has systematically applied the relativistic 
Hartree Fock method to calculate RE transition probabilities. Even with an 
intense theoretical effort on these complex spectra, it is essential to have 
some good measurements for comparison. Fortunately an efficient experimental 
approach has been developed. Many of the recent experimental studies of RE 
atomic transition probabilities have combined radiative lifetimes from laser 
induced fluorescence (LIF) measurements with emission branching fractions 
measured using a Fourier transform spectrometer (FTS). This approach to 
determining atomic transition probabilities in complex spectra has proved to 
be both efficient and quite reliable. Our recent work on Sm \textsc{ii} is 
just one example, and many other studies on RE transition probabilities are 
cited in the first paragraph of our Sm \textsc{ii} paper (Lawler et al. 
2006, hereafter LDSC06). 

Singly-ionized gadolinium is one of the remaining RE species in need of 
modern measurements, and it is the focus of this work. We report new LIF 
radiative lifetime measurements for 49 even-parity and 14 odd-parity levels, 
as well as absolute atomic transition probabilities for 611 lines of Gd 
\textsc{ii}. Our lifetime measurements are in good agreement with earlier, 
but less extensive, LIF measurements. Our branching fraction measurements, 
in combination with the LIF lifetimes, yielded both a large set of 
transition probabilities and the first based on modern methods. This 
improved data set has been used to determine a new solar photospheric Gd 
abundance and revised Gd abundances for the $r-$process-rich metal-poor giant 
stars CS 22892-052, BD+17$^{o}$3248, and HD 115444. Implications of these 
abundance determinations are discussed.

\begin{center}
2. RADIATIVE LIFETIME MEASUREMENTS
\end{center}

Radiative lifetimes of 49 even-parity and 14 odd-parity levels of Gd 
\textsc{ii} have been measured using time-resolved laser-induced 
fluorescence (LIF) on an atom/ion beam. Only a cursory description of the 
experimental method is given here, since the apparatus and technique have 
been described in many previous publications on other species. The reader is 
referred to recent work in Eu \textsc{i}, \textsc{ii}, and \textsc{iii} (Den 
Hartog et al. 2002) for a more detailed description.

A slow ($\sim $5$\times $10$^{4}$~cm/s), weakly collimated beam of Gd atoms 
and ions is produced using a hollow cathode discharge sputter source. A 
pulsed argon discharge, operating at $\sim $0.4~torr with 10~$\mu $s 
duration, 10~A pulses, is used to sputter the gadolinium which lines the 
hollow cathode. The hollow cathode is closed on one end except for a 1~mm 
hole, through which the gadolinium atoms and ions flow into a low pressure 
(10$^{-4}$~torr) scattering chamber. This beam is intersected at right 
angles by a nitrogen laser-pumped dye laser beam 1~cm below the cathode 
bottom. The laser is tunable over the range 2050 - 7200~{\AA} with the use 
of frequency doubling crystals, is pulsed at $\sim $30~Hz repetition rate 
with a $\sim $3~ns pulse duration, and has a $\sim $0.2~cm$^{-1}$ bandwidth. 
The laser is used to selectively excite the level to be studied, eliminating 
the possibility of cascade radiation from higher-lying levels. 

Fluorescence is collected at right angles to the laser and atomic/ionic 
beams through a pair of fused-silica lenses which form an f/1 optical 
system, and detected with a RCA 1P28A photomultiplier tube (PMT). Optical 
filters, either broadband colored glass filters or narrowband multi-layer 
dielectric filters, are typically inserted between the two lenses to cut 
down on scattered laser light and to block cascade radiation from lower 
levels. The signal from the PMT is recorded and averaged over 640 shots 
using a Tektronix SCD1000 digitizer. Data collection begins after the laser 
pulse has terminated to make deconvolution of the laser excitation 
unnecessary. Data are recorded with the laser tuned on and off the 
excitation transition. A linear least-square fit to a single exponential is 
performed on the background-subtracted fluorescence decay to yield the 
lifetime of the level. The lifetime is measured twice for each level, using 
a different excitation transition whenever possible. This redundancy helps 
ensure that the transitions are identified correctly in the experiment, 
classified correctly and are free from blends.

The lifetimes reported here have an uncertainty of $\pm $5{\%}, except for 
the shortest lifetimes ($<$4 ns) for which the uncertainties are $\pm $0.2 
ns. To achieve this level of fidelity and maintain it over the full dynamic 
range of the experiment (2~ns to 1.5~$\mu $s for ions), the possible 
systematic errors in these measurements must be well understood and 
controlled. They include electronic bandwidth limitations, cascade 
fluorescence, Zeeman quantum beats and atomic motion flight-out-of-view 
effects, among others. The dominant systematic error depends on the 
lifetime, for example the bandwidth, linearity, and overall fidelity of the 
electronic detection system prevents us from achieving better than $\pm $0.2 
ns accuracy even on short lifetimes. These systematic effects are discussed 
in detail in earlier publications, (See, for example, Den~Hartog et al. 
1999; 2002) and will not be discussed further here. As a means of verifying 
that the measurements are within the stated uncertainties, we perform 
periodic end-to-end tests of the experiment by measuring a set of well known 
lifetimes. These cross-checks include lifetimes of Be \textsc{i} (Weiss 
1995), Be \textsc{ii} (Yan et al. 1998) and Fe \textsc{ii} (Guo et al. 1992; 
Bi\'{e}mont et al. 1991), covering the range from 1.8--8.8 ns. An Ar~ 
\textsc{i} lifetime is measured at 27.85~ns (Volz {\&} Schmoranzer 1998). He 
\textsc{i} lifetimes are measured in the range 95 -- 220~ns (Kono {\&} 
Hattori 1984). 

The results of our lifetime measurements of 49 even-parity and 14 odd-parity 
levels of Gd \textsc{ii} are presented in Table 1. Energy levels are from 
the tabulation by Martin et al. (1978). Air wavelengths are calculated from 
the energy levels using the standard index of air (Edl\'{e}n 1953). The 
uncertainty of the lifetimes is the larger of $\pm $5{\%} or $\pm $0.2 ns.

Also presented in Table~1 is a comparison of our results with those from 
other LIF lifetime measurements available in the literature. We see very 
good agreement with the recent work of Xu et al. (2003). They measured 
lifetimes for 13 even-parity levels of Gd \textsc{ii}, of which 11 were in 
common with our study. All of our measurements agreed within our joint error 
bars except for the level at 29198 cm$^{-1}$, which was only slightly worse 
than that. We found a mean difference between our measurements and theirs of 
0.0{\%} and a rms difference of 5.6{\%}. Our measurements also agree very 
well with those of Zhang et al. (2001a). They measured lifetimes for 20 
even-parity levels of Gd \textsc{ii}, all of which overlap with our study. 
All of the lifetimes agreed within the joint uncertainties and we see a 
--0.4{\%} mean difference and 6.8{\%} rms difference between our 
measurements and theirs. The comparison to the three lifetimes measured by 
Bergstr\"{o}m et al. (1988) is less satisfactory, particularly for the level 
30009 cm$^{-1}$, for which their result is considerably lower than our 
measurement and that of Zhang et al.. The mean and rms differences between 
our measurements are --15.8{\%} and 17.9{\%}, respectively, for these three 
levels.

Two older sets of measurements exist giving lifetimes of a total of six 
levels of Gd \textsc{ii}, using the delayed-coincidence method with 
non-selective electron beam excitation. Gorshkov et al. (1983) report 
lifetimes on four levels and Gorshkov {\&} Komarovskii (1986) on two 
additional. These lifetimes are all substantially longer than ours and other 
LIF results (by as much as a factor of 3), probably due to cascade from 
higher-lying levels because of the non-selective nature of the electron beam 
excitation. These results are not included in Table 1.

\begin{center}
3. BRANCHING FRACTIONS AND ATOMIC TRANSITION PROBABILITIES
\end{center}

Branching fraction measurements in complex RE spectra such as Gd \textsc{ii} 
require an extremely powerful spectrometer. As in earlier work on RE 
spectra, we used the 1.0 meter FTS at the National Solar Observatory (NSO) 
for this project. This instrument has the large etendue of all 
interferometric spectrometers, a limit of resolution as small as 0.01 
cm$^{-1}$, wavenumber accuracy to 1 part in 10$^{8}$, broad spectral 
coverage from the UV to IR, and the capability of recording a million point 
spectrum in 10 minutes (Brault 1976). An FTS is insensitive to any small 
drift in source intensity since an interferogram is a simultaneous 
measurement of all spectral lines.

Figure 1 is a partial Grotrian diagram for Gd \textsc{ii}. Although there 
are nine valence electrons in singly ionized Gd, seven of the nine form a 
half completed 4f shell. The high spin of the half completed 4f shell leads 
to rich, but not overwhelmingly complex, energy level structure. Low odd- 
and even-parity configurations are built by putting the remaining two 
valence electrons into some combination of 5d, 6s, 4f, and/or 6p orbitals. 
Most of the low odd- and even-parity levels have sufficiently pure LS 
coupling, that they can be assigned using that coupling scheme. Energy 
ranges of important low-lying odd- and even-parity levels of each 
sub-configuration are marked in Fig. 1, as well as the energy ranges and 
sub-configurations of the upper levels studied in this work.

The nine valence electrons of singly ionized Gd yield low odd-parity levels, 
including the ground level, in the 4f$^{7}(^{8}$S)(5d+6s)$^{2}$ 
sub-configurations. The tight coupling of the 4f electrons to form the 
$^{8}$S parent term is not broken until about 32,500 cm$^{-1 }$(Cowan 1981, 
Blaise et al. 1971). All 20 levels of the $^{10}$D$^{o}$, $^{8}$D$^{o}$, 
$^{8}$D$^{o}$, and $^{6}$D$^{o}$ terms of the 4f$^{7}(^{8}$S)5d6s 
sub-configuration are known, as is the 4f$^{7}(^{8}$S)6s$^{2 8}$S$^{o}$ 
level. The 4f$^{7}(^{8}$S)5d$^{2}$ sub-configuration contains 43 levels in 
the $^{8}$G$^{o}$, $^{10}$F$^{o}$, $^{8}$F$^{o}$, $^{6}$F$^{o}$, 
$^{8}$D$^{o}$, $^{10}$P$^{o}$, $^{8}$P$^{o}$, $^{6}$P$^{o}$, and 
$^{8}$S$^{o}$ terms, all of which are also known. These low odd-parity 
levels are spread from the ground 4f$^{7}(^{8}$S)5d($^{9}$D)6s 
$^{10}$D$^{o}_{5/2 }$level at 0.00 cm$^{-1}$ to the 
4f$^{7}(^{8}$S)5d$^{2}(^{1}$S) $^{8}$S$^{o}_{7/2 }$level at $\sim 
$29,000 cm$^{-1}$. The next band of observed odd-parity levels are part of 
the 4f$^{8}(^{7}$F)6p$^{ }$sub-configuration that starts at $\sim $ 33,000 
cm$^{-1}$ (see Table IX of Blaise et al. 1971 for predictions of unobserved 
4f$^{8}$6p$^{ }$ levels). 

The low even-parity levels start just under 8000 cm$^{-1}$ with 13 levels of 
the $^{8}$F and $^{6}$F terms of the 4f$^{8}(^{7}$F)6s$^{ 
}$sub-configuration. The 57 levels of the $^{8}$H, $^{6}$H, $^{8}$G, 
$^{6}$G, $^{8}$F, $^{6}$F, $^{8}$D, $^{6}$D, $^{8}$P, and $^{6}$P terms of 
the 4f$^{8}(^{7}$F)5d$^{ }$sub-configuration are all known. These levels 
start $\sim $18,000 cm$^{-1}$ and extend to $\sim $32,000 cm$^{-1}$. 
Although the 4f$^{7}(^{8}$S)(6s+5d)6p$^{ }$ sub-configuration starts around 
26,000 cm$^{-1}$, it appears that all even-parity levels in the 25,000 
cm$^{-1}$ to 30000 cm$^{-1}$ range have been observed (see Fig. 1 of Blaise 
et al. 1971). Two perturbed levels assigned to the 4f$^{8}(^{5}$D)6s$^{ 
}$ sub-configuration have been found at 26352 cm$^{-1}$ and 27274 
cm$^{-1}$(Blaise et al. 1971). The absence of low-lying unobserved levels 
simplified the assessment of missing branches or residuals in our branching 
fraction measurements and provided high confidence in a partition function 
evaluation. The interleaving of low even- and odd-parity levels made it 
possible for us to measure lifetimes and transition probabilities for upper 
levels of both parities.

The upper levels studied in this work have significant 6p character as 
indicated in Table 1. The odd-parity upper levels are part of the 
4f$^{8}(^{7}$F)6p sub-configuration. The even-parity upper levels are from 
the 4f$^{7}(^{8}$S)6s6p and 4f$^{7}(^{8}$S)5d6p sub-configurations. 
Energy ranges of the upper levels are indicated in Fig. 1.

In order to make our branching fraction measurements as complete as 
possible, we worked on the 16 spectra listed in Table 2. We recorded some of 
these spectra during observing runs in the 2000 through 2002 period and 
extracted the older spectra from the NSO electronic archives\footnote{ The 
NSO archives are available at http://nsokp.nso.edu/dataarch.html}. The 
former spectra are of commercially manufactured, sealed Gd hollow cathode 
discharge (HCD) lamps with fused silica windows containing either argon or 
neon buffer gas fills. Although these small lamps typically yield spectra 
with minimal optical depth or radiation trapping effects, we still recorded 
a few spectra at reduced currents to check for such effects. Our most useful 
spectra were recorded using the small lamps at 20 to 30 mA currents with 
many, $\sim $ 50, coadds. These small commercial lamps were designed for 
atomic absorption spectrophotometers used by analytical chemists, and they 
are usually very stable which is crucial for recording good interferograms. 
(Lamp oscillations can easily wreck an interferogram by introducing spurious 
modulations that result in ghosts in the final spectrum.) The 20 to 30 mA 
currents are above the manufacturer's recommended maximum current. We used 
forced air cooling during all-night integrations to get our most useful 
spectra. This approach shortens the lamp lifetimes, but yields the most 
valuable, optically thin spectra with good S/N ratios covering wide spectral 
regions. The custom (water cooled) HCD lamp and electrodeless discharge lamp 
(EDL) yielded spectra with superior S/N ratios for many weak lines with 
branching fractions as small as 0.001. The high currents ($>$ 100 mA) in the 
custom HCD resulted in some optical depth effects on the strongest Gd 
\textsc{ii} lines to low odd-parity levels. These potential errors were 
identified and eliminated by comparing the high- and low-current HCD 
spectra. The comparison of spectra with Ne and Ar buffer gas was used to 
eliminate potential errors from blends of buffer gas lines with lines of Gd 
\textsc{ii}.

The establishment of an accurate relative radiometric calibration or 
efficiency is critical to a branching fraction experiment. As indicated in 
Table 2, we made greater use of standard lamp calibrations in this Gd 
\textsc{ii} study than in previous RE studies. We are constantly trying to 
improve our radiometric calibrations of the FTS, because such calibrations 
are thought to be the dominant source of uncertainty for many of our final 
log(\textit{gf}) values. Tungsten (W) filament standard lamps are particularly useful 
near the Si detector cutoff in the 10,000 to 9,000 cm$^{-1}$ range where the 
FTS sensitivity is changing rapidly as a function of wave number, and near 
the dip in sensitivity at 12,500 cm$^{-1}$ from the aluminum coated optics. 
Tungsten lamps are not bright enough to be useful for FTS calibrations in 
the UV region, and UV branches typically dominate the decay of levels 
studied using our lifetime experiment. In general one must be careful when 
using continuum lamps to calibrate the FTS over wide spectral ranges, 
because the ``ghost'' of a continuum is a continuum. The Ar \textsc{i} and 
Ar \textsc{ii} line technique, which is internal to the HCD Gd/Ar lamp 
spectra, is still our preferred calibration technique. It captures the 
wavelength-dependent response of detectors, spectrometer optics, lamp 
windows, and any other components in the light path or any reflections which 
contribute to the detected signal (such as due to light reflecting off the 
back of the hollow cathode). This calibration technique is based on a 
comparison of well-known branching ratios for sets of Ar \textsc{i} and Ar 
\textsc{ii} lines widely separated in wavelength, to the intensities 
measured for the same lines. Sets of Ar \textsc{i} and Ar \textsc{ii} lines 
have been established for this purpose in the range of 4300 to 35000 
cm$^{-1}$ by Adams {\&} Whaling (1981), Danzmann {\&} Kock (1982), 
Hashiguchi {\&} Hasikuni (1985), and Whaling et al. (1993). One of our best 
Gd/Ar HCD spectra from 2002, and some the Gd/Ar HCD spectra from 1991, can 
also be calibrated with tungsten standard lamp spectra recorded shortly 
before, or after, the HCD lamp spectra. The older tungsten lamp is a strip 
lamp calibrated as a spectral radiance (W/(m$^{2}$ sr nm)) standard, and the 
newer is a tungsten-quartz-halogen lamp calibrated as a spectral irradiance 
(W/(m$^{2}$ nm) at a specified distance) standard. Neither of these filament 
lamps is hot or bright enough to yield a reliable UV calibration, but they 
are useful in the visible and near IR for interpolation and as a redundant 
calibration. Argon calibration lines were largely absent in the EDL spectra, 
and thus these spectra were calibrated using the tungsten lamps.

All possible transition wave numbers between known energy levels of Gd 
\textsc{ii} satisfying both the parity change and $\Delta $J = -1, 0, or 1 
selection rules were computed and used during analysis of FTS data. Energy 
levels from Martin et al. (1978) were used to determine possible transition 
wave numbers. Levels from Martin et al. (1978) are available in electronic 
form from Martin et al. (2000)\footnote{Available at 
http://physics.nist.gov/cgi{\-}bin/AtData/main{\_}asd}.

Branching fraction measurements were attempted on lines from all 63 levels 
of the lifetime experiment, and were completed for lines from 39 even-parity 
and 13 odd-parity upper levels. The levels for which branching fractions 
could not be completed had a strong branch beyond the UV limit of our 
spectra, or had a strong branch which was severely blended. Typically an 
upper level, depending on its J value, has about 30 possible transitions to 
known lower levels. More than 20,000 possible spectral line observations 
were studied during the analysis of 16 different Gd/Ar and Gd/Ne spectra. We 
set baselines and integration limits ``interactively'' during analysis of 
the FTS spectra. The same numerical integration routine was used to 
determine the un-calibrated intensities of Gd \textsc{ii} lines and selected 
Ar \textsc{i} and Ar \textsc{ii} lines used to establish a relative 
radiometric calibration of the spectra. A simple numerical integration 
technique was used in this and most of our other RE studies because of 
weakly resolved or unresolved hyperfine and isotopic structure. More 
sophisticated profile fitting is used only when the line sub-component 
structure is either fully resolved in the FTS data or known from independent 
measurements.

The procedure for determining branching fraction uncertainties was described 
in detail by Wickliffe et al. (2000). Branching fractions from a given upper 
level are defined to sum to unity, thus a dominant line from an upper level 
has small branching fraction uncertainty almost by definition. Branching 
fractions for weaker lines near the dominant line(s) tend to have 
uncertainties limited by their S/N ratios. Systematic uncertainties in the 
radiometric calibration are typically the most serious source of uncertainty 
for widely separated lines from a common upper level. We used a formula for 
estimating this systematic uncertainty that was presented and tested 
extensively by Wickliffe et al. (2000). The spectra of the high current HCD 
lamp and EDL's enabled us to connect the stronger visible and near IR 
branches to quite weak branches in the same spectral range. Uncertainties 
grew to some extent from piecing together branching ratios from so many 
spectra, but such effects have been included in the uncertainties on 
branching fractions of the weak visible and near IR lines. In the final 
analysis, the branching fraction uncertainties are primarily systematic. 
Redundant measurements with independent radiometric calibrations help in the 
assessment of systematic uncertainties. Redundant measurements from spectra 
with different discharge conditions also make it easier to spot blended 
lines and optically thick lines. 

Branching fractions from the FTS spectra were combined with the radiative 
lifetime measurements described in {\S}2 to determine absolute transition 
probabilities for 611 lines of Gd \textsc{ii} in Table 3. Air wavelengths in 
Table 3 were computed from energy levels (Martin et al. 1978) using the 
standard index of air (Edl\'{e}n 1953). Parities are included in Table 3 
using the ``ev'' and ``od'' notation introduced in Table 1. 

Transition probabilities for the very weakest lines (branching fractions $<$ 
0.001) which were observed with poor S/N ratios and for a few blended lines 
are not included in Table 3, however these lines are included in the 
branching fraction normalization. The effect of the problem lines becomes 
apparent if one sums all transition probabilities in Table 3 from a chosen 
upper level, and compares the sum to the inverse of the upper level lifetime 
from Table 1. Typically the sum of the Table 3 transition probabilities is 
between 95{\%} and 100 {\%} of the inverse lifetime. Although there is 
significant fractional uncertainty in the branching fractions for these 
problem lines, this does not have much effect on the uncertainty of the 
stronger lines that were kept in Table 3. Branching fraction uncertainties 
are combined in quadrature with lifetime uncertainties to determine the 
transition probability uncertainties in Table 3. Possible systematic errors 
from missing branches to unknown lower levels are negligible in Table 3, 
because we were able to make at least rough measurements on visible and near 
IR lines with branching fractions as small as 0.001 . The radiative 
lifetimes of the Gd \textsc{ii} levels in this study are generally shorter 
than the radiative lifetimes we studied in Sm \textsc{ii} (LDSC06). The 
generally short Gd \textsc{ii} lifetimes, in combination with the frequency 
cubed scaling of transition probabilities, means that any unknown line in 
the mid- to far-IR region will not have a significant branching fraction. 

We have searched the literature unsuccessfully for any recent branching 
fraction measurements or calculations on Gd \textsc{ii}. The only published 
branching fraction or transition probability measurements we found are based 
on photographic data from the National Bureau of Standards. Relative 
intensity measurements by Meggers et al. (1961) were converted to absolute 
transition probabilities by Corliss {\&} Bozman (1962). Ward (1985) reported 
a formula for re-normalizing the Corliss {\&} Bozman (1962) transition 
probabilities. Cowley {\&} Corliss (1983) developed a formula for 
determining transition probabilities from line intensities published by 
Meggers et al. (1975) which are an updated version of the original Meggers 
et al. (1961) line intensities used by Corliss {\&} Bozman (1962). The 
problems with the 1961 data set are illustrated by efforts to renormalize 
it, and they have been discussed extensively (e.g. Obbarius {\&} Kock 1982). 

\begin{center}
4. SOLAR AND STELLAR GADOLINIUM ABUNDANCES
\end{center}

We have employed the new Gd \textsc{ii} transition probabilities to 
re-determine gadolinium abundances for the solar photosphere and three very 
metal-poor ([Fe/H] $<$ -2)\footnote{ We adopt standard stellar spectroscopic 
notations that for elements A and B, [A/B] = 
log$_{10}$(N$_{A}$/N$_{B})_{star}$ - log$_{10}$(N$_{A}$/N$_{B})_{sun}$, 
for abundances relative to solar, and log $\varepsilon $(A) = 
log$_{10}$(N$_{A}$/N$_{H})$ + 12.0, for absolute abundances.} stars that 
have large overabundances of the rare-earth elements. These are stars 
enriched in rapid $n-$capture ($r-$process) nucleosynthesis products: HD 115444 
([Fe/H] = -2.9, [Eu/Fe] = +0.8, Westin et al. 2000); BD+17$^{o}$3248 ([Fe/H] 
= -2.1, [Eu/Fe] = +0.9, Cowan et al. 2002), and CS 22892-052 ([Fe/H] = -3.1, 
[Eu/Fe] = +1.5, Sneden et al. 2003). Our abundance study followed the 
methods used in previous papers of this series, most closely resembling 
those employed for Sm \textsc{ii} by LDSC06.

\bigskip
4.1 Line Selection
\bigskip

We have accurate transition probabilities for 611 Gd \textsc{ii} lines, but 
not all of these lines are useful in Gd abundance analyses. For our program 
objects, the majority of the Gd \textsc{ii} lines prove to be either 
undetectably weak, or heavily blended with other species transitions, or 
both. Here we describe the Gd \textsc{ii} line selection procedures. As 
discussed by LDSC06, in a standard LTE abundance analysis the relative 
strengths of lines of an individual species vary directly as the transition 
probabilities modified by the Boltzmann excitation factors. Thus for a weak 
line on the linear part of the curve-of-growth the equivalent width (EW) and 
reduced width (RW) are related as, log(RW) $\equiv $ log(EW/$\lambda )$ 
$\propto $ log(\textit{gf}) - $\theta \chi $, where excitation energy $\chi $ is in 
units of eV and inverse temperature $\theta  \quad \equiv $ 5040/T. The relative 
strengths of lines of different species also depend on relative elemental 
abundances and Saha ionization equilibrium factors. However, Gd and Sm have 
similar low ionization potentials, 6.150 eV and 5.644 eV, respectively 
(Grigoriev {\&} Melikhov 1997). (All RE's have first ionization potentials 
within 0.5 eV of 5.9 eV.) In line-forming atmospheric layers ($\tau  \quad \ge $ 
0.1) of the Sun and stars considered here, Gd and Sm exist almost 
exclusively in their ionized states: n$_{II}$/n$_{I} \quad >$ 100, or n$_{II 
}\approx $ n$_{total}$. Therefore, for the ionized-species transitions 
studied by LDSC06 and here, the Saha corrections to account for other 
ionization state populations are negligible. In this case the relative 
strength factors of weak lines can be written as STR $\equiv $ 
log($\varepsilon $\textit{gf}) $-- \theta \chi $,  
where $\varepsilon $ is the elemental 
abundance.

In Figure 2 we plot these relative strength factors as a function of 
wavelength for Sm \textsc{ii} (LDSC06) and Gd \textsc{ii} lines. To compute 
these strength factors we have adopted solar abundances of log $\varepsilon 
$(Sm) = +1.00 (LDSC06), and log $\varepsilon $(Gd) = +1.10 (close to the 
recommended photospheric abundance of Grevesse {\&} Sauval 2002, Lodders 
2003, and the new value derived in this paper). 

In Figure 2 we have indicated the approximate minimum strength factor for 
lines at the detection threshold in the solar spectrum. This limiting 
strength case has been computed as follows. LDSC06 searched the very 
high-resolution, high S/N solar center-of-disk spectrum of Delbouille et al. 
(1973) for the weakest solar Sm \textsc{ii} lines that could be reliably 
detected and employed in an abundance analysis. That exercise suggested a 
lower limit for unblended lines of EW $\approx $ 1.5 m{\AA} in the blue 
spectral region ($\lambda  \quad \sim $ 4500 {\AA}), or log(RW) $\approx $ -6.5. 
Lines of Sm \textsc{ii} near this limit had values of log(\textit{gf}) -- $\theta 
\chi  \quad \approx $ -1.6, which translates to STR $\approx $ -0.6. The 
equivalent width detection limit should also apply to Gd \textsc{ii}, and so 
that limit has been indicated in both panels of the figure with horizontal 
dotted lines. 

In Figure 2 we also show the minimum strength factor for Sm \textsc{ii} and 
Gd \textsc{ii} lines that exhibit substantial absorption in the solar 
spectrum. This is a fairly arbitrary assignment. Beginning at the defined 
detection-limit STR = -0.6, a line 20 times stronger will have STR = -0.6 + 
1.3 = +0.7. If such lines remained on the linear (unsaturated) part of the 
curve-of-growth then the increase in equivalent width would be identical: 
log(RW) = -6.5 + 1.3 = -5.2, or EW $\approx $ 30 m{\AA} near 4500 {\AA}. In 
reality lines in this RW regime are slightly saturated; test calculations 
suggest that the solar Sm \textsc{ii} and Gd \textsc{ii} lines with strength 
factors 20 times larger than the detection limit will have log(RW) $\approx 
$ -5.35, or EW $\approx $ 20 m{\AA} at 4500 {\AA}. We adopt STR = +0.7 as 
the lower limit for strong lines in this study, and have drawn dashed 
horizontal lines to indicate this in the figure.

Inspection of Figure 2 reveals similarities and differences in the 
transitions of Sm \textsc{ii} and Gd \textsc{ii}. First, essentially all 
useful lines for a solar abundance analysis have wavelengths $\lambda  \quad <$ 
5000 {\AA}. This is especially true for Gd \textsc{ii}, for which nearly all 
lines above the defined detection threshold occur at $\lambda  \quad <$ 4500 
{\AA}. Second, Gd \textsc{ii} has many more strong lines (more than 40 with 
STR $>$ +0.7) potentially available for abundance analysis than did Sm 
\textsc{ii} (only 4). However, all strong lines of both species are located 
in the near-UV, $\lambda  \quad <$ 4000 {\AA}, where the line density is large. 
None of these lines are unblended, so that attempts to derive abundances 
from an equivalent width analysis are risky. For Gd \textsc{ii} especially, 
the lack of many potentially detectable lines in the less crowded spectral 
regions redward of 4500 {\AA} emphasizes the necessity of computing full 
synthetic spectra for all transitions used in the final analysis.

As in LDSC06, the strength factor plot of Figure 2 was used to make the 
first cut in reducing the original large list of Gd \textsc{ii} lines to a 
more manageable list for the solar/stellar work. Of the 611 lines with 
laboratory transition probabilities newly determined in this paper, 235 have 
STR $\ge $ -0.6. Attempts to detect useful transitions among the 376 Gd 
\textsc{ii} lines below this strength level in the solar spectrum failed 
with one exception (see below). These weaker lines were therefore discarded.

We then followed the procedures described in LDSC06 and earlier papers of 
this series to identify the final set of Gd \textsc{ii} lines to be used in 
the solar/stellar abundance analyses. With the aid of the Delbouille et al. 
(1973) solar center-of-disk spectrum, the Moore, Minnaert, {\&} Houtgast 
(1966) solar line identification atlas, the Kurucz (1998) atomic and 
molecular line compendium, and the observed spectrum of the $r-$process-rich 
metal-poor giant star BD+17\r{ }3248 (Cowan et al. 2002), we eliminated 
about 150 more Gd \textsc{ii} lines that proved to be undetectably weak, 
extremely blended, or both. Some examples of this elimination process are 
discussed by LDSC06. All of the preliminary culling efforts finally produced 
a list of about 80 lines worthy of closer inspection in solar and/or stellar 
spectra. 

\bigskip
4.2 The Solar Photospheric Gadolinium Abundance
\bigskip

We employed synthetic spectrum computations to determine a new gadolinium 
abundance for the solar photosphere. The procedures were identical to those 
described by LDSC06. We employed Kurucz's (1998) line database and Moore et 
al.'s (1966) solar identifications to generate lists of relevant atomic and 
molecular lines in 4-6 {\AA} regions surrounding each Gd \textsc{ii} 
transition. We used (a) these line lists, (b) the Holweger {\&} M\"{u}ller 
(1974) solar model atmosphere, and (c) a standard solar abundance set drawn 
from reviews by Grevesse {\&} Sauval (1998, 2002) and Lodders (2003) 
supplemented by values determined in earlier papers of this series, as 
inputs into the current version of the LTE line analysis code MOOG (Sneden 
1973) to generate the synthetic spectra. We adopted some well-determined 
transition probabilities for ionized species of neutron-capture elements 
from the following sources: Gd, the present work; La, Lawler et al. (2001a); 
Nd, Den Hartog et al. (2003); Eu, Lawler et al. (2001b); Sm, LDSC06, Tb, 
Lawler et al. (2001c); Dy, Wickliffe et al. (2000); Ho, Lawler et al. 
(2004); Ce, Palmeri et al. (2000); Y, Hannaford et al. (1982); and Zr, 
Malcheva et al. (2006).

We computed multiple synthetic spectra for each Gd line region, and compared 
them to the Delbouille et al. (1973) center-of-disk photospheric spectrum. 
We smoothed the spectra empirically by a Gaussian to match the observed line 
broadening due to solar macroturbulence and spectrograph instrumental 
effects. The oscillator strengths for contaminant atomic transitions (except 
for the species listed above) were adjusted to fit the solar spectrum. 
Molecular line strengths were altered as a group via abundance changes of C, 
N, or O as appropriate. For unidentified solar features, we arbitrarily 
added Fe \textsc{i} lines with excitation potentials $\chi $ = 3.5 eV to the 
line lists. The initial synthetic spectrum computations showed that the 
majority of the proposed Gd \textsc{ii} transitions are very blended or very 
weak in the solar spectrum, thus useless in a Gd photospheric abundance 
analysis. The iterated line lists for these discarded solar Gd \textsc{ii} 
features were retained for further investigation in the $r-$process-rich stellar 
spectra ({\S} 4.3).

In the end we used just 20 carefully selected Gd \textsc{ii} lines in the 
solar spectrum. In the left-hand panels (a), (b), and (c) of Figure 3 we 
show synthetic and observed spectra of three representative transitions. 
Panel (a) contains two neighboring Gd \textsc{ii} lines that are commonly 
employed in studies of $r-$process-rich metal-poor stars because they are 
relatively strong and located in the commonly observed spectral region near 
4000 {\AA}. Panel (b) shows a relatively strong line (STR $\approx $ +1.1) 
that suffers only modest contamination from other species in spite of its 
location at 3549 {\AA}. Finally, panel (c) demonstrates that one of the 
strongest Gd \textsc{ii} lines (STR $\approx $ +1.3) can be detected with 
confidence even though its wavelength of 3358 {\AA} lies in the middle of a 
strong band of NH. 

The derived abundances for individual Gd\textsc{ ii} lines are listed in 
column 4 of Table 4, and are displayed as a function of wavelength in Figure 
4. A straight mean abundance is log $\varepsilon $(Gd) = 1.11 $\pm $ 0.01 
($\sigma $ = 0.05, 20 lines). The abundances show no obvious trend with 
wavelength (Figure 4), excitation potential (although the range is small), 
log(\textit{gf}), or general line strength. 

The two lines at the wavelength extremes of our solar list deserve comment. 
Moore et al. (1966) identifies a solar absorption at 5733.89 {\AA} as Gd 
\textsc{ii}, with EW = 1 m{\AA}, or log(RW) = -6.8. This is the weakest line 
that they attribute to Gd \textsc{ii}, and is 0.3 dex smaller than our 
suggested weak-line limit. The absorption at this wavelength appears also in 
the Delbouille et al. (1973) photospheric spectrum, so we included it in our 
analysis, deriving log $\varepsilon $ = 1.15, consistent with the mean 
abundance. We were unable to identify any other solar Gd lines in this 
strength regime. Moore et al. also identify a strong line at 3032.84 {\AA} 
as Gd \textsc{ii}. Because this line is in the very crowded near-UV spectral 
region and their identifications were from relatively noisy photographic 
spectra, Moore et al. did not estimate an equivalent width for this line. 
Our analysis confirms its identification, and the derived abundance of log 
$\varepsilon $ = 1.10 is in excellent accord with the mean. The 3032 {\AA} 
line is the shortest-wavelength solar feature that we have been able to 
model successfully in this series of papers.

Abundance uncertainties can be due to line profile matching factors 
(internal uncertainties) and scale factors (external uncertainties). For the 
present Gd analysis these issues are nearly identical to those outlined for 
Sm by LDSC06. Transition profile fitting uncertainties are estimated at $\pm 
$0.02 dex, and on average the uncertainties due to contamination by other 
species lines are also $\pm $0.02 dex. The mean error in log(\textit{gf}) for the 20 
lines used in the solar analysis (see Table 4) is $\pm $0.03. Adding these 
uncertainties in quadrature yields an estimated total internal uncertainty 
of $\pm $0.04 dex, which is close to the observed $\sigma $ = 0.05. 

Overall scale errors can arise from other atomic data uncertainties and 
model atmosphere choices. As stated in {\S}4.1, Saha-fraction corrections 
are negligible for Gd \textsc{ii}, so the derived Gd abundance depends 
directly on the Boltzmann factor and the Gd \textsc{ii} partition function. 
We checked Irwin's (1981) polynomial representation of the temperature 
dependent Gd \textsc{ii} partition function against a partition function 
evaluated from the online NIST energy levels (Martin et al. 2000). We found 
nearly perfect agreement, as expected, because the lower levels of Gd 
\textsc{ii} which determine the partition function for T $<$ 6000 K are all 
known. 

The influence of solar model atmosphere choice was assessed by repeating 
sample abundance derivations using the Kurucz (1998) and Grevesse {\&} 
Sauval (1999) models, finding on average abundance shifts of -0.01 and -0.02 
dex, respectively, with respect to those derived with the Holweger {\&} 
M\"{u}ller (1974) model. These differences are nearly identical to those 
determined for other rare earth ions in the previous papers of this series. 
Therefore abundance scale errors appear to be very small, of order 0.02 dex, 
within the limits imposed by our analysis assumptions (single-stream, 
plane-parallel atmosphere geometry, LTE). Combining internal line-to-line 
scatter uncertainties (which contribute just $\pm $0.01 in the mean 
abundance, since 20 lines are employed here) and external scale 
uncertainties, we recommend log $\varepsilon $(Gd)$_{Sun}$ = +1.11 $\pm $ 
0.03.

This new Gd abundance is in agreement with the result of the only other 
solar photospheric analysis in the past two decades. From an equivalent 
width analysis of eight lines, Bergstr\"{o}m et al. (1988) derived log 
$\varepsilon $(Gd)$_{Sun}$ = +1.12 $\pm $ 0.04, where their uncertainty 
estimate was set to twice the standard deviation of the mean abundance. Note 
that seven of their chosen eight lines have also been included in our study. 
The photospheric abundance appears to be slightly higher than the 
recommended meteoritic abundances in two recent compilations: log 
$\varepsilon $(Gd)$_{met}$ = +1.06 $\pm $ 0.02 (Lodders 2003), and +1.03 
$\pm $ 0.02 (Asplund, Grevesse, {\&} Sauval 2005) . This point will be 
considered again in {\S}5.

\bigskip
4.3 Gadolinium Abundances in Three $r-$Process-Rich Low Metallicity Stars
\bigskip

We next explored the Gd \textsc{ii} spectra of very metal-poor, 
$r-$process-rich giant stars BD+17$^{o}$3248, HD 115444, and CS 22892-052. 
Spectra of such stars present much more favorable cases for the study of Gd 
\textsc{ii} and other rare-earth first ions. This is due to the confluence 
of several effects: overall metal deficiency; relative $n-$capture-element 
enhancement; and lower stellar temperatures and gravities than the Sun 
(which combine to weaken the numerous high excitation neutral-species 
transitions and strengthen low excitation ionized-species ones). Many Gd 
\textsc{ii} lines that are too blended and/or weak in the solar spectrum can 
be analyzed reliably in the $r-$process-rich stars. We illustrate this in 
right-hand panels (d), (e), and (f) of Figure 3, displaying the same Gd 
\textsc{ii} lines that were previously shown in the solar spectrum. The 
central depths of the Gd lines are similar in the Sun and BD+17$^{o}$3248. 
However, the total absorptions and thus equivalent widths are larger in 
BD+17$^{o}$3248 because the line breadths are larger in the star (this is 
mainly an effect of the coarser resolution of this and the other two stellar 
spectra). Although the Gd \textsc{ii} lines are typically stronger in the 
$r-$process-rich stars than in the Sun, the lower resolution and S/N of the 
stellar spectra changed the detection-with-confidence limit to about 3 
m{\AA} near 4500 {\AA}, about double that of the solar spectrum. Therefore 
the strength factor detection limit was roughly the same in solar and 
stellar cases, and we made no attempt to re-visit the entire Gd line list to 
discover additional very weak Gd \textsc{ii} lines.

We derived Gd abundances for the stars in the same manner as was described 
for the Sun in {\S} 4.2. The abundances from individual lines are listed in 
Table 4 and displayed in Figure 4. The mean abundances, standard deviations, 
and number of lines are given at the bottom of Table 4 and Figure 4. The 
line-to-line scatters are all small, $\sigma $ = 0.04 - 0.07, and are mainly 
due to stellar spectrum measurement uncertainties. 

\begin{center}
5. ABUNDANCES OF $n-$CAPTURE ELEMENTS IN METAL-POOR HALO STARS
\end{center}

The new laboratory atomic data, particularly transition probabilities, are 
critical to 

abundance determinations of the $n-$capture elements in the metal-poor (old) 
Galactic halo stars. These ongoing abundance studies are providing new 
information about the nature of the earliest Galactic nucleosynthesis and 
the nature of the earliest stars - those that preceded the formation of the 
halo stars (Cowan {\&} Sneden 2006). Increasingly more accurate stellar 
abundances have also recently allowed detailed comparisons with solar system 
distributions (Den Hartog et al. 2003; Lawler et al. 2004, 2006). Such 
abundance comparisons are providing new and more complete understandings of 
the nuclear processes and the astrophysical sites for heavy element 
nucleosynthesis.

In Figure 5 we illustrate the $n-$capture abundances in the atomic number range 
56 $\le $ Z $\le $ 68 for the Solar System and for the three very metal-poor 
([Fe/H] $<$ -2) halo giant stars CS 22892-052, HD 115444 and BD+17\r{ }3248. 
We have plotted the abundance differences (log observed abundance \textit{minus} log solar 
system $r-$process only value) for each element in each star. For this 
comparison we have normalized the abundance distributions of all three stars 
at the $r-$process element Eu. The solar system elemental 

$r-$process abundance distribution was obtained by summing the individual 
$r-$process isotopic abundance contributions, based upon the so-called standard 
model (see Simmerer et al. 2004; Cowan et al. 2006 and the discussion 
below).

If the stellar and solar $r-$process abundance values were identical they would 
fall on the solid horizontal line in the figure -- i.e., log $\varepsilon 
$(X)$_{obs}$ - log $\varepsilon $(X)$_{s.s.(r-only)}$ = 0. Abundance 
comparisons between the stellar elemental abundances for Gd and other 
$n-$capture elements with the total Solar System meteoritic abundance values 
(dotted-line curve) from Lodders (2003) are also shown. We show the 
abundances in the top panel of this figure from the original published 
papers by our group (Westin et al. 2000; Cowan et al. 2002; Sneden et al. 
2003). It is clear that there was a large amount of scatter for a number of 
elements including Nd, Sm, Ho and Gd in the stellar data. Reducing or 
eliminating this abundance scatter has been one of the prime motivations to 
obtain improved laboratory data for various elements of astrophysical 
interest.

In the bottom panel of Figure 5 we show the newly revised abundances, 
utilizing the new transition probabilities for the elements Nd (Den Hartog 
et al. 2003), Ho (Lawler et al. 2004), Sm (LDSC06) and Gd from this paper. 
As a result of employing these new atomic data, the star-to-star scatter is 
greatly reduced, and there is good agreement between the elemental values in 
CS 22892-052, HD 115444 and BD+17\r{ }3248 and the solar system $r-$process 
values. This good agreement is further support for the finding that the 
abundances of the stable elements (at least for Ba and above) are consistent 
with the scaled solar system elemental $r-$process distribution. (see, $e.g.$, Truran 
et al. 2002; Sneden {\&} Cowan 2003 and Cowan {\&} Sneden 2006). It also 
again demonstrates that early in the history of the Galaxy the $r-$process was 
the dominant synthesis mechanism, as the $n-$capture elements seen in these 
stars were formed in the $r-$process only, and not the $s-$process. Many of the rare 
earth elements have a significant $s-$process component in solar system material 
($e.g.$, Nd is 58{\%} $s-$process and Sm is 33{\%} $s-$process, Simmerer et al. 2004) -- 
this can be seen in Figure 5 as the differences between the total solar 
photospheric abundances (black dots with dotted line from Lodders 2003) and 
the stellar ($r-$process) elemental abundances. However, since the predominant 
$s-$process synthesis is coming from low-mass long-lived stars (Busso et al. 
1999), there is not sufficient time early in the history of the Galaxy for 
these stars to live, die and eject $s-$process enriched material into the ISM 
prior to the formation of the observed halo stars. Instead, all of these 
elements must have been synthesized in relatively high-mass, 
rapidly-evolving stars that presumably exploded as core-collapse supernovae 
and enriched the gas in the early Galaxy (Cowan {\&} Thielemann 2004; Cowan 
{\&} Sneden 2006). 

\bigskip
5.1 Solar and Stellar Abundance Comparisons
\bigskip

The detailed solar and stellar abundance comparisons depend upon the 
deconvolution of Solar System material into separate isotopic $s-$ and 
$r-$process contributions. The individual $s-$process isotopic abundances are first 
determined 

(often) employing the so called ``classical model'' approximation in 
conjunction with measured neutron capture cross-sections. A more complicated 
``stellar model'' approach to obtaining the $s-$process contributions has also 
been made (Arlandini et al. 1999). 

Cross-section measurements are not possible for the far more radioactive and 
short-lived $r-$process nuclei. The $r-$process isotopic abundances (or residuals) 
are therefore determined by just subtracting the calculated $s-$process 
abundances from the total Solar System abundances. We have tabulated in 
Cowan et al. (2006) the individual $s-$ and $r-$process isotopic solar system 
abundances (based upon the Si = 10$^{6}$ scale and assuming the classical 
model approximation) from the work of K\"{a}ppeler et al. (1989), Wisshak et 
al.(1998) and O'Brien et al. (2003). The solar system $r-$process (only) 
elemental abundance curve, such as the one employed in Figure 5 (and labeled 
as log $\varepsilon $(X)$_{S.S.(r-only)}$, is obtained by summing the 
individual isotopic contributions from the $r-$process. (Similarly, the 
$s-$process only elemental abundance curve is the sum of the $s-$process 
contributions.)

In spite of Figure 5's very good overall agreement between the abundances of 
the rare-earth elements in the halo stars and the solar system $r-$process 
abundances discussed above, some small deviations have become more apparent 
as RE transition probabilities have improved. For example, in the bottom 
panel of Figure 5 the Gd abundances in the three halo stars are clustered 
together slightly above the (meteoritic-based) Solar System $r-$process (only) 
value -- implying that the solar system abundance might be too low. 

The increasingly accurate stellar abundances could possibly be used to 
predict specific $r-$process abundances directly, rather than obtaining the 
residuals in the manner described above. We have attempted to do this, based 
upon the following procedure. First, we list in Table 5 the $s-$process 
(N$_{s})$ and $r-$process (N$_{r})$ contributions to a RE element. Those 
abundances (based upon the Si = 10$^{6}$ scale and assuming the classical 
model approximation) are listed in Table 5 (see also Cowan et al. 2006 for a 
complete list). Next we determined the difference between log(N$_{r}$(el)) - 
log(N$_{r}$(Eu)) for each element. That result was compared with the average 
difference between the RE elements and Eu (almost entirely an $r-$process 
element), $<$el-Eu$>$, for the three halo stars. Assuming that this average 
value was the correct one, we obtained a predicted solar system $r-$process 
abundance, N$_{r}$(predicted) -- this would be the value that would raise 
the solid line to be coincident with the stellar data. Thus, in the case of 
Gd, we obtain a value of N$_{r}$(predicted) = 0.312, rather than the 
previously determined value of 0.276. This implies that the solar 
$r-$process only value for Gd should be raised by that difference. Finally, 
assuming that the previously predicted $s-$process elemental contribution is 
correct, we have then obtained the total solar system abundances for those 
elements by summing N$_{s}$ and N$_{r}$(predicted). Those values, along with 
log $\varepsilon _{total}$, are also listed in Table 5.

While it is clear that we have made several simplifying assumptions in 
making these new predictions, several points are worth noting. In the case 
of Gd we find that the stellar data suggests a higher recommended value for 
the $r-$process contribution and total Solar System value. Interestingly, this 
total predicted value for Gd, log $\varepsilon _{total}$ = 1.11, is 
identical to our new measured photospheric abundance for this element. For 
Sm our predicted value, log $\varepsilon _{total}$ = 1.00, is also 
identical to the measured solar photospheric value (LDSC06) and for Ho our 
prediction, 0.53, is consistent with recent measured photospheric values log 
$\varepsilon _{total}$ = 0.51 $\pm $ 0.1 (Lawler et al. 2004) and log 
$\varepsilon _{total}$ = 0.53 $\pm  \quad \approx $ 0.1 (Bord {\&} Cowley 
2002). In the case of Nd, we find very little difference between the older 
predicted values and the new one based upon the stellar data. Our predicted 
solar abundance, log $\varepsilon _{total}$ = 1.46, is in very good 
agreement with both the meteoritic and photospheric values, including the 
recent photospheric measurement of log $\varepsilon _{total}$ = 1.45 $\pm 
$ 0.01 by Den Hartog et al. (2003) and the recommendation of Lodders (2003) 
of 1.46 based upon the meteoritic measurements. Our analyses may suggest, at 
least for the cases of Gd, Sm and Ho, that the (slightly) higher 
photospheric determinations (including recent ones found by our group) might 
be the recommended solar values, or at least be more appropriate for stellar 
abundance studies, than the meteoritic ones.

We caution in this numerical analysis, however, that there are some 
inconsistencies in the data sources regarding the $s-$process contributions, 
which in turn effects the $r-$process residual abundances. Many of the 
$s-$process determinations, for instance, were obtained using older cross 
section measurements (K\"{a}ppeler et al. 1989) and older Solar System 
abundance determinations (Anders {\&} Grevesse 1984). The specific 
$s-$process (and $r-$process) contributions are based upon individual isotopic 
neutron capture cross sections, and assume a specific $s-$process model - we 
have chosen to employ the ``standard model''. Elemental abundance values are 
the then sum of the isotopic values. Since there are some uncertainties in 
the cross sections and in the assumed model predictions, these elemental 
sums sometime can have slight deviations from the predicted total solar 
values. In addition since that time, some of the cross sections (and hence 
$s-$process contributions) have been updated, as has the total Solar System 
abundance predictions (Lodders 2003). Thus, our precise (total) $r-$process 
abundance numerical values predicted for the Solar System have to be viewed 
with some caution, or at least with some error bars. Nevertheless, we think 
the general procedure is sound -- that the halo stars abundances can be used 
to predict (or at least constrain) the Solar System $r-$process values -- and 
the general trends suggest a revision may be needed in those current values. 
Clearly, a new systematic analysis of the $s-$process contributions (based upon 
new nuclear cross section experiments) needs to be performed on the latest 
solar system abundance determinations (Lodders 2003).

Many of other possible sources of systematic error were discussed in earlier 
sections. Most of these effects will shift the absolute $r-$process scale 
without changing the internal $r-$process pattern. The similarity of RE 
excitation and ionization potentials tends to ``cancel out'' effects from 
choosing a slightly different photospheric model or slightly different model 
parameters. The use of many metal lines in each abundance determination 
suppresses errors from blending and continuum placement during analysis of 
stellar data. The overall ``scale'' uncertainties on the laboratory 
transition probability data from the LIF radiative lifetime measurements are 
quite small as described in {\S}2 and illustrated in LDSC06 with a more 
extensive comparison of independent sets of LIF measurements on Sm 
\textsc{ii}. If the difficult radiometric calibrations of the FTS data 
needed for branching fraction measurements were seriously in error, then we 
would expect to see a wavelength dependence in the elemental abundances. 
Figure 4 is a crucial test for this systematic problem. Incomplete knowledge 
of RE energy levels and level assignments was a significant concern during 
our work on Sm \textsc{ii}, but it was not a problem in the present work on 
Gd \textsc{ii}. This is one systematic which could change the internal 
$r-$process pattern. If there were significant residuals, or long wavelength 
transitions to unobserved lower levels in Sm \textsc{ii}, then the log(\textit{gf}) 
values used in the Sm \textsc{ii} abundance determinations would need to be 
decreased and the Sm abundances would increase. The net result would move 
the Sm abundances further above the solid line of Figure 5, which is not the 
direction needed to support the conventional $r-$process abundance pattern 
determined by subtracting calculated $s-$process abundances from the solar 
(photospheric) or solar system (meteoric) abundance pattern. 

\bigskip
5.2 Isotopic Considerations for Gd \textsc{ii}
\bigskip

Gadolinium has seven abundant naturally occurring isotopes: $^{152}$Gd 
(0.2{\%} of the solar-system elemental abundance), $^{154}$Gd (2.2{\%}), 
$^{155}$Gd (14.8{\%}), $^{156}$Gd (20.5{\%}), $^{157}$Gd (15.6{\%}), 
$^{158}$Gd (24.8{\%}), and $^{160}$Gd (21.9{\%}). We have listed the values 
from both the standard and stellar models (Arlandini et al. 1999) in Table 
6. The abundances for the $s-$ and $r-$process contributions are based upon the Si = 
10$^{6}$ scale (see Cowan et al. 2006). We have also listed the percentage 
contribution by individual isotope to the total elemental $s-$ and $r-$process 
abundances (i.e., the vertical columns add up to 100{\%} in those particular 
columns). It is clear from the table that both the standard and stellar 
models give very similar isotopic abundance predictions for each of the $s-$ and 
$r-$process mixtures for Gd.

The widths of line profiles in our high resolution FTS data vary, and in a 
few cases the profiles have partially resolved structure. Although it is not 
possible today, it may at some point in the future be possible to observe 
the isotopic mixture of Gd in a metal-poor halo star, similarly to what 
Lambert {\&} Allende Prieto (2002) have done for the element Ba in the halo 
star HD 140283. They determined that the fractional abundance of the odd 
isotopes of Ba 

$f_{odd}$ = [N($^{135}$Ba) + N($^{137}$Ba)]/N(Ba)

in this star was consistent with the solar system $r-$process isotopic ratio. Gd 
is an even-Z nucleus (like Ba and Sm) and has seven stable ($s-$ and $r-$process 
admixed) isotopes. Thus, (similarly to Ba and Sm) we can define for Gd 

$f_{odd}$ = [N($^{155}$Gd) + N($^{157}$Gd)]/N(Gd).

For the pure $r-$process components of solar system isotopic abundances, we find 
that $f^{ r}_{odd}$ = 0.33 employing values from either the standard model 
(Cowan et al. 2006) or the stellar model (Arlandini et al. 1999) that are 
listed in Table 6. Interestingly, this result for Gd is almost identical to 
the value (0.36) found for Sm (LDSC06). (Note that this result and the 
values listed in Table 6, come from the original prediction for the total 
$r-$process contribution to Gd, N$_{r}$, and not on the new suggested value 
discussed above.)

For comparison, the Gd solar $s-$process values are $f^{ s}_{odd}$ = 0.16 and 
$f^{ s}_{odd}$ = 0.17 for the standard and stellar models, respectively. 

There have been very few stellar isotopic abundance observations. These 
include only the one measurement of Ba in HD 140283 by Lambert and 
Allende-Prieto and several Eu isotopic observations (Sneden et al. 2002, 
Aoki et al. 2003). An observation of the isotopic mixture for Gd in any halo 
star, would give important information about early Galactic nucleosynthesis. 
For example, such an observation of Gd (82{\%} $r-$process), perhaps in 
conjunction with another element such as Sm (67{\%} $r-$process in solar system 
material) in the same star,would provide a direct measure of the $r-$process 
contribution to the elemental Gd (and perhaps Sm) production in 
nucleosynthetic (e.g., supernovae) sites that were operating in the early 
Galaxy. Further, it would confirm that not only the elemental, but the 
isotopic abundances of these elements are consistent with the Solar System 
$r-$process distribution. Such observations would also be a measure of the 
``robustness'' of the $r-$process (operating in approximately the same manner 
over many Gyr between the formation of the Galaxy and the Solar System) and 
a confirmation that the Solar System abundances are in many ways cosmic.

\newpage
\begin{center}
6. SUMMARY AND CONCLUSIONS
\end{center}

New LIF radiative lifetime measurements for 63 levels of Gd \textsc{ii} and 
FTS branching fraction/atomic transition probability measurements for 611 
transitions of Gd \textsc{ii} were completed and reported herein. These 
laboratory measurements were applied to re-determine the Solar photospheric 
abundance of Gd, and to extend our effort to more sharply define a pure 
$r-$process abundance pattern using the metal-poor Galactic halo stars CS 
22892-052, BD+17$^{o}$3248, and HD 115444. A sharply defined $r-$process 
abundance pattern will provide a strong constraint for advanced models of 
this process. 

We have employed the increasingly accurate stellar abundance determinations, 
resulting in large part from the more precise laboratory atomic data, to 
predict directly the Solar System $r-$process elemental abundances for Gd, Sm, 
Ho and Nd. Our analysis of the stellar data suggests slightly higher 
recommended values for the $r-$process contribution and total Solar System 
values. These values are consistent with recent photospheric determinations 
(including those from our group), for the elements Gd, Sm and Ho. This may 
suggest that these slightly higher photospheric values might be the 
recommended solar values, or at least be more appropriate for stellar 
abundance studies. Similarly to the case of Sm, we have analyzed the 
isotopic mixture of Gd providing some odd-isotope, and $r-$process ratio, 
predictions that could be utilized in future Gd isotopic studies. The 
combination of improved laboratory data, better observational data, and 
advanced models will unambiguously identify the site(s) of the $r-$process, and 
fully elucidate the role of the $r-$process in the chemical evolution of the 
Galaxy and the Universe.

\begin{center}
ACKNOWLEDGMENTS
\end{center}
This work is supported by the National Science Foundation under grants AST- 
0506324 (JEL and EADH), AST{\-}0307495 (CS), and AST-0307279 (JJC). J. E. 
Lawler is a guest observer at the National Solar Observatory and he is 
indebted to Mike Dulick and Detrick Branstron for help with the 1 m Fourier 
transform spectrometer. We thank K. Lodders for helpful comments about the 
Solar System abundances.

\begin{center}
REFERENCES
\end{center}

Adams, D. L., {\&} Whaling, W. 1981, J. Opt. Soc. Am., 71, 1036 

Aoki, W., Ryan, S. G., Iwamoto, N., Beers, T. C., Norris, J. E., Ando, H., 
Kajino, T., Mathews, G. J., {\&} Fujimoto, M. Y. 2003, ApJ, 592, L67

Arlandini, C., K\"{a}ppeler, F., Wisshak, K., Gallino, R., Lugaro, M., 
Busso, M., {\&} Straniero, O. 1999, ApJ, 525, 886

Asplund, M., Grevesse, N., {\&} Sauval, A. J. 2005, ASP Conf. Ser. 336: 
\textit{Cosmic 
Abundances as Records of Stellar Evolution and Nucleosynthesis}, 336, 25

Bergstr\"{o}m, H., Bi\'{e}mont, E., Lundberg, H., Persson, A. 1988, A{\&}A, 
192, 335.

Bi\'{e}mont, E., Baudoux, M., Kurucz, R. L., Ansbacher, W., {\&} Pinnington, 
E. H. 1991, 
A{\&}A, 249, 539

Bi\'{e}mont, E. {\&} Quinet P. 2003, Physica Scripta, T105, 38

Blaise, J., van Kleef, Th. A. M., Wyart, J. F. 1971, J. de Phys., 32, 617 

Bord, D. J. {\&} Cowley, C. R. 2002, Sol. Phys., 211, 3

Brault, J. W. 1976, J. Opt. Soc. Am., 66, 1081

Busso, M., Gallino, R., {\&} Wasserburg, G. J. 1999, ARA{\&}A, 37, 239

Cayrel, R., et al.,  2004, A{\&}A, 416, 1117

Corliss, C. H., {\&} Bozman, W. R. 1962, \textit{Experimental Transition Probabilities for }
\textit{Spectral Lines of Seventy Elements}, U. S. Natl. Bur. Standards Monograph 53,
(Washington: U. S. Government Printing Office)

Cowan, J. J., Burris, D. L., Sneden, C., McWilliam, A., {\&} Preston, G. W. 
1995, ApJ, 
439, L51

Cowan, J. J., Sneden, C., Burles, S., Ivans, I. I., Beers, T. C., Truran, J. 
W., Lawler, J. E.,
Primas, F., Fuller, G. M., Pfeiffer, B., {\&} Kratz, K.-L. 2002, ApJ,. 572, 
861

Cowan, J. J., {\&} Thielemann, F.-K. 2004, Phys. Today, 57, 47

Cowan, J. J., Lawler, J. E., Sneden, C., Den Hartog, E. A., {\&} Collier, J. 
2006, To appear
in Proc. 2006 NASA LAW ed. V. Kwong

Cowan, J. J., {\&} Sneden, C. 2006, Nature, 440, 1151

Cowan, R. D. 1981, \textit{The Theory of Atomic Structure and Spectra} (Berkeley, Univ. of
California Press) p 601

Cowley, C. R. {\&} Corliss, C. H. 1983, MNRAS 203, 651

Danzmann, K., {\&} Kock, M. 1982, J. Opt. Soc. Am., 72, 1556

Delbouille, L, Roland, G., {\&} Neven, L. 1973, 
\textit{Photometric Atlas of the Solar Spectrum 
from $\lambda $ 3000 to $\lambda $ 10000}, (Li\`{e}ge, Inst. D'Ap., Univ. de Li\`{e}ge)

Den Hartog, E. A., Lawler, J. E., Sneden, C., {\&} Cowan, J. J. 2003, ApJS, 
148, 543

Den~Hartog, E. A., Wickliffe, M. E., {\&} Lawler, J. E. 2002, ApJS, 141, 255

Den~Hartog, E. A., Wiese, L. M. {\&} Lawler, J. E. 1999, J. Opt. Soc. Am. B, 
16, 2278

Edl\'{e}n, B. 1953, J. Opt. Soc. Am., 43, 339

Gorshkov, V. N., {\&} Komarovskii, V. A. 1986, Sov. Astron. 30, 333 [orig. 
1986, Astron.
Zh., 63, 563], 

Gorshkov, V. N., Komarovskii, V. A., Osherovich, A. L., {\&} Penkin, N. P. 
1983, Opt. 
Spectrosc. (USSR), 54, 122 [orig. 1983, Opt. Spektrosk. 54, 210]

Gratton, R. G., {\&} Sneden, C. 1994, A{\&}Ap, 287, 927

Grevesse, N., {\&} Sauval, A. J. 1998, Space Sci. Rev., 85, 161

Grevesse, N., {\&} Sauval, A. J. 1999, A{\&}A, 347, 348

Grevesse, N., {\&} Sauval, A. J. 2002, Adv. Space. Res., 30, 3

Grigoriev, I. S., {\&} Melikhov, E. Z. 1997, 
\textit{Handbook of Physical Quantities}, (Boca Raton,
CRC Press) p. 516

Guo, B., Ansbacher, W., Pinnington, E. H., Ji, Q., {\&} Berends, R. W. 1992, 
Phys. Rev. A, 
46, 641

Hannaford, P., Lowe, R. M., Grevesse, N., Biemont, E., {\&} Whaling, W. 
1982, ApJ, 
261, 736

Hashiguchi, S., {\&} Hasikuni, M. 1985, J. Phys. Soc. Japan 54, 1290

Holweger, H., {\&} M\"{u}ller, E. A. 1974, Sol. Phys., 39, 19

Irwin, A. W. 1981, ApJS, 45, 621

K\"{a}ppeler, F., Beer, H., {\&} Wisshak, K. 1989, Rep. Prog. Phys., 52, 945

Kono, A., {\&} Hattori, S. 1984, Phys. Rev. A, 29, 2981

Kurucz, R. L. 1998, in \textit{Fundamental Stellar Properties: The Interaction between 
Observation and Theory}, IAU Symp. 189, ed T. R. Bedding, A. J. Booth and J. 
Davis (Dordrecht: Kluwer), p. 217

Lambert, D.L., {\&} Allende Prieto, C. 2002, MNRAS, 335, 325

Lawler, J. E., Bonvallet, G., {\&} Sneden, C. 2001a, ApJ, 556, 452

Lawler, J. E., Wickliffe, M. E., Den Hartog, E. A., {\&} Sneden, C. 2001b, 
ApJ, 563, 1075

Lawler, J. E., Wickliffe, M. E., Cowley, C. R., {\&} Sneden, C. 2001c, ApJS, 
137, 341

Lawler, J. E., Sneden, C., {\&} Cowan, J. J. 2004, ApJ, 604, 850

Lawler, J. E., Den Hartog, E. A., Sneden, C., {\&} Cowan, J. J. 2006, ApJS, 
162, 227

Lodders, K. 2003, ApJ, 591, 1220

Malcheva, G., Blagoev, K., Mayo, R., Ortiz, M., Xu, H. L., Svanberg, S., 
Quinet, P., {\&} Bi\'{e}mont, E. 2006, MNRAS, 367, 754

Martin, W.C., Zalubas, R., {\&} Hagan, L. 1978,\textit{ Atomic Energy Levels {\-} The Rare Earth Elements}, NSRDS{\-}NBS 60 
(Washington: U. S. G. P. O.) p. 174

Martin, W. C., Sugar, J., {\&} Musgrove, A. 2000, NIST Atomic Spectra 
Database,
(http://physics.nist.gov/cgi{\-}bin/AtData/main{\_}asd)

McWilliam, A., Preston, G. W., Sneden, C., {\&} Searle, L. 1995, AJ, 109, 
2757

Meggers, W. F., Corliss, C. H., and Scribner, B. F. 1961, \textit{Tables of Spectral Line Intensities}, U. S. Natl. Bur. 
Standards Monograph 32, (Washington: U.S. G.P.O.)

Meggers, W. F., Corliss, C. H., and Scribner, B. F. 1975, \textit{Tables of Spectral Line}
\textit{Intensities}, U. S. Natl. Bur. Standards Monograph 145, (Washington: U.S. G. P.
O.)

Moore, C. E., Minnaert, M. G. J., {\&} Houtgast, J. 1966, \textit{The Solar Spectrum 2934 {\AA} to }
\textit{8770 {\AA}}, NBS Monograph 61 (Washington: U.S. G. P. O.)

Obbarius H. U., {\&} Kock, M. 1982, J. Phys. B: At. Mol. Opt. Phys., 15, 527

O'Brien, S., Dababneh, S., Heil, M., K\"{a}ppeler, F., Plag, R., Reifarth, 
R., Gallino. R., {\&} Pignatari, M. 2003, Phys. Rev. C, 68, 035801

Palmeri, P., Quinet, P., Wyart, J.-F., {\&} Bi\'{e}mont, E. 2000, Physica 
Scripta, 61, 323

Ryan, S. G., Norris, J. E., and Beers, T. C. 1996, ApJ, 471, 254

Simmerer, J., Sneden, C., Cowan, J. J., Collier, J., Woolf, V. M., {\&} 
Lawler, J. E. 2004,
ApJ, 617, 1091

Sneden, C. 1973, ApJ, 184, 839

Sneden, C., et al.,  
2003, ApJ, 591,  936

Sneden, C., McWilliam, A., Preston, G. W., Cowan, J. J., Burris, D. L., {\&} 
Armosky, B. 
J. 1996, ApJ,  467, 819

Sneden, C., Cowan, J. J., Lawler, J. E., Burles, S., Beers, T. C., Fuller, 
G. M. 2002, ApJ 566, L25

Sneden, C., {\&} Cowan, J. J. 2003, Science, 299, 70

Truran, J. W., Cowan, J. J., Pilachowski, C. A., {\&} Sneden, C. 2002, PASP, 
114, 1293

Volz, U., {\&} Schmoranzer, H. 1998, in AIP Conf. Proc. 434, \textit{Atomic and 
Molecular Data and Their Applications}, ed. P. J. Mohr and W. L. 
Wiese (Woodbury, NY:AIP), p. 67

Ward, L. 1985, Mon. Not. R. Astr. Soc. 213, 17

Weiss, A. W. 1995, Phys. Rev. A, 51, 1067

Westin, J., Sneden, C., Gustafsson, B., {\&} Cowan, J.J. 2000, ApJ, 530, 783

Whaling, W., Carle, M. T., {\&} Pitt, M. L. 1993, J. Quant. Spectrosc. 
Radiat. Transfer 50, 7

Wickliffe, M. E., Lawler, J. E., {\&} Nave, G. 2000, J. Quant. Spectrosc. 
Radiat. Transfer,  66, 363

Wisshak, K., Voss, F., K\"{a}ppeler, F., Kazakov, L., {\&} Reffo, G. 1998, 
Phys. Rev. C, 57, 391

Xu, H., Jiang, Z., {\&} Svanberg, S. 2003, J. Phys. B, 36, 411 

Yan, Z-C, Tambasco, M., {\&} Drake, G. W. F. 1998, Phys. Rev. A, 57, 1652

Zhang, Z. G., Persson, A., Li, Z. S., Svanberg, S., {\&} Zhankui, J. 2001a, 
Eur. Phys. J. D,  13, 301

\newpage 
\begin{center}
FIGURE CAPTIONS
\end{center}
\bigskip
Figure 1: Partial Grotrian diagram for singly ionized Gd.

\bigskip

Figure 2: Relative transition strength factors, STR $\equiv $ 
log($\varepsilon $\textit{gf}) -- $\theta \chi $, for lines of Sm \textsc{ii} 
(LDSC06) and Gd \textsc{ii} (this study). For display purposes the 
long-wavelength limit has been set to 8000 {\AA}, which cuts out only some 
extremely weak lines of Gd \textsc{ii} and Sm \textsc{ii} that can be 
detected neither in the Sun nor nearly all other stars. The short-wavelength 
limit of 2900 {\AA} covers all lines at that end of the spectrum in these 
two studies. Definitions of ``detection limit'' and ``strong lines'' of 
these species are given in the text.

\bigskip

Figure 3: Representative Gd \textsc{ii} lines at 4037.32 and 4037.89 {\AA} 
(top panels (a) and (d)), 3549.36 {\AA} (middle panels (b) and (e)), and 
3358.63 {\AA} (bottom panels (c) and (f)) in spectra of the Sun (left-hand 
panels (a), (b), and (c)) and the $r-$process-rich metal-poor giant 
BD+17$^{o}$3248 (right-hand panels (d), (e), and (f)). In each panel there 
are four synthetic spectra drawn with solid lines. The synthetic spectrum 
with the weakest (sometimes totally absent) Gd \textsc{ii} line was computed 
assuming no Gd contribution to the total absorption. For the other three 
syntheses, the middle-strength one was computed with the best-fit abundance 
to the Gd feature given in Table 4 (except for the 4037 {\AA} pair, where 
the best compromise abundance between the two lines was used). The 
stronger/weaker syntheses surrounding the best-fit one were done assuming Gd 
abundances that were a factor of two larger/smaller. The filled circles 
represent the observed spectra. In the solar case, for display purposes we 
chose to plot the Delbouille et al. (1973) data at intervals of 0.01 {\AA} 
instead of the original 0.002 {\AA}.

\bigskip

Figure 4: Line-by-line Gd abundances for the Sun (x symbols) and the 
$r-$process-rich metal-poor giant stars BD+17$^{o}$3248 (open circles), CS 
22892-052 (plus signs), and HD 115444 (diamonds), plotted as a function of 
wavelength. For display purposes the long-wavelength end of the plot has 
been truncated at 4700 {\AA}. This cuts just one line from the plot, that at 
5733 {\AA}, which was detected only in the solar spectrum. For each star, a 
dotted line is drawn at the mean abundance. As indicated in the figure 
legend, the three numbers in parentheses beside each star name are the mean 
abundance, the sample standard deviation $\sigma $, and the number of lines 
used in the analysis.

\bigskip

Figure 5. Neutron-capture elemental abundance patterns in CS 22892-052, HD 
115444, and BD+17\r{ }3248 compared with the (scaled) Solar System 
$r-$process abundances (solid line) and the total Solar System meteoritic 
abundances recommended by Lodders (2003; dashed line). This figure is an 
update of Figure 13 in LDSC06. The abundance
distributions of all of the stars have been normalized to that of Eu. In the 
top panel, stellar abundances are those reported in the original papers on 
these stars. In the bottom panel the comparisions are repeated for these 
stars, but substituting in the abundances of Nd, Ho, and Sm, and now Gd 
derived in papers of this series. Also shown are solar photospheric 
abundances, with values of La, Nd, Sm, Eu, Gd, Tb, and Ho taken from this 
series, otherwise from Lodders (2003).

\end{document}

\tablenum{2}
\begin{deluxetable}{|c|c|c|c|c|c|c|c|c|c|c|c|c|}
\tabletypesize{\scriptsize}
\rotate
\tablewidth{0pt}
\tablecaption{
Fourier transform spectra of Gd lamps used in this study. All were 
recorded using the 1 m FTS on the McMath telescope at the National Solar 
Observatory, Kitt Peak, AZ. }
%\begin{center} 

%\begin{longtable}[-3.5in]
%{|p{18pt}|p{30pt}|p{17pt}|p{35pt}|l|l|l|l|p{20pt}|l|l|p{35pt}|p{40pt}|}
%\hline
%\endhead
%\hline
%\endfoot
\tablehead{
Index& 
Date& 
Serial& 
Lamp& 
Buffer& 
Lamp& 
Wavenumber& 
Limit of& 
Coadds& 
Beam& 
Filter& 
Detector$^{2}$& 
Calibration$^{3}$ \\
%\hline
& 
& 
No.& 
Type$^{1}$& 
Gas& 
Current& 
Range& 
Resolution& 
& 
Splitter& 
& 
& 
 \\
%\hline
& 
& 
& 
& 
& 
(mA)& 
(cm$^{-1})$& 
(cm$^{-1})$& 
& 
& 
& 
& 
}
% \\
%\hline
\startdata
1& 
2002 Feb. 27& 
18& 
Commercial  \par HCD& 
Ar& 
26& 
7929 -- 34998& 
0.050& 
50& 
UV& 
& 
S. B. Si Diode& 
Ar I {\&} II  WQH Lamp \\
\hline
2& 
2000 Feb. 28& 
23& 
Commercial  \par HCD& 
Ar& 
23& 
7929 -- 34998& 
0.053& 
16& 
UV& 
& 
S. B. Si Diode& 
Ar I {\&} II \\
\hline
3& 
2000 Feb. 27& 
18& 
Commercial  \par HCD& 
Ar& 
30& 
7929 -- 34998& 
0.053& 
41& 
UV& 
& 
S. B. Si Diode& 
Ar I {\&} II \\
\hline
4& 
2001 Feb. 27& 
19& 
Commercial  \par HCD& 
Ar& 
15& 
7929 -- 34998& 
0.053& 
9& 
UV& 
& 
S. B. Si Diode& 
Ar I {\&} II \\
\hline
5& 
2002 Feb. 27& 
20& 
Commercial  \par HCD& 
Ar& 
12.5& 
7929 -- 34998& 
0.053& 
9& 
UV& 
& 
S. B. Si Diode& 
Ar I {\&} II \\
\hline
6& 
1991 Oct. 10& 
13& 
Custom HCD& 
Ar& 
295& 
15154 - 36081& 
0.044& 
4& 
UV& 
CuSO$_{4}$& 
S. B. Si Diode& 
Ar I {\&} II  W Strip Lamp \\
\hline
7& 
1991 Oct. 10& 
14& 
Custom HCD& 
Ar& 
290& 
15154 - 36081& 
0.044& 
4& 
UV& 
CuSO$_{4}$& 
S. B. Si Diode& 
Ar I {\&} II  W Strip Lamp \\
\hline
8& 
1991 Oct. 9& 
11& 
Custom HCD& 
Ar& 
300& 
7810 -- 25033& 
0.031& 
4& 
UV& 
GG 495& 
S. B. Si Diode& 
W Strip Lamp \\
\hline
9& 
1991 Oct. 9& 
12& 
Custom HCD& 
Ar& 
300& 
7810 -- 25033& 
0.031& 
4& 
UV& 
GG 495& 
S. B. Si Diode& 
W Strip Lamp \\
\hline
10& 
1991 Dec. 12& 
7& 
Custom HCD& 
Ar& 
300& 
1661 -- 11312& 
0.015& 
8& 
CaF$_{2}$& 
GaAs& 
InSb& 
Ar I {\&} II  W Strip Lamp \\
\hline
11& 
1991 Dec. 12& 
8& 
Custom HCD& 
Ne& 
300& 
1661 -- 11312& 
0.015& 
8& 
CaF$_{2}$& 
GaAs& 
InSb& 
W Strip Lamp \\
\hline
12& 
2001 Jan. 25& 
13& 
Commercial  \par HCD& 
Ne& 
25& 
7929 - 34998& 
0.050& 
10& 
UV& 
& 
S. B. Si Diode& 
 \\
\hline
13& 
2001 Jan. 26& 
15& 
Commercial  \par HCD& 
Ne& 
20& 
7929 -- 34998& 
0.050& 
10& 
UV& 
& 
S. B. Si Diode& 
 \\
\hline
14& 
1985 Feb. 6& 
10& 
EDL& 
Ar& 
& 
7456 -- 28808& 
0.035& 
2& 
UV& 
GG 375& 
S. B. Si Diode& 
W Strip Lamp \\
\hline
15& 
1985 Feb. 6& 
67& 
EDL& 
Ar& 
& 
7456 -- 28808& 
0.035& 
8& 
UV& 
GG 375& 
S. B. Si Diode& 
W Strip Lamp \\
\hline
16& 
1985 Feb. 6& 
89& 
EDL& 
Ar& 
& 
3285 -- 15050& 
0.018& 
8& 
UV& 
RG 715& 
InSb& 
W Strip Lamp
\enddata
%\label{tab1}
%\end{longtable}
%\end{center}
\par\noindent
$^{1}$Lamp types include commercially available small sealed Hollow 
Cathode Discharge (HCD) lamps typically used in atomic absorption 
spectrophotometers, a custom water-cooled HCD lamp,  
and an 
Electrodeless Discharge Lamp (EDL). \\
%\par\noindent
%\newline
%\setlength{\parindent}{-20mm}
$^{2}$Detectors types include the Super Blue (S. B.) Si photodiode.
\par\noindent
$^{3}$Relative radiometric calibrations were based on selected sets 
of Ar I and Ar II lines, on a Tungsten-Quartz-Halogen (WQH) lamp 
calibrated as a secondary irradiance standard, and on  
a 
Tungsten (W) Strip Lamp calibrated as a secondary radiance standard.
\end{deluxetable}